\documentclass[10pt,aps.prl,twocolumn,superscriptaddress]{revtex4-1} %% REVTeX 4.0
\usepackage[usenames]{color}

\usepackage{graphics}
\usepackage{epsfig}
\usepackage{amsmath}
\usepackage{float}
\usepackage{subfig}
\usepackage{tabularx}
\usepackage[normalem]{ulem}
\usepackage{color}

\begin{document}
\title{A White Light Interferometric Microscope for Measuring Dose-Dependent Reversible Photodegradation}
\author{Benjamin Anderson, Elizabeth Bernhardt and  Mark G. Kuzyk}
\address{Department of Physics and Astronomy, Washington State University,
Pullman, WA 99164-2814}
\date{\today}

\begin{abstract}
We have developed a white light interferometric microscope (WLIM), which can spatially resolve the change in the complex index of refraction, and apply it to study reversible photodegradation of 1,4-Diamino-9,10-anthraquinone doped into PMMA.  The measured change in absorbance is consistent with standard spectrometer measurements. \textcolor{black}{The refractive index of a pristine sample measured with the WLIM is also found to be consistent with the spectrum found by independent means.\cite{Yakuphanoglu08.01} We report on measurements of the change in refractive index due to photodegradation, which is found to be consistent with Embaye's two-population model\cite{embaye08.01}.}  We show the WLIM can be used as a powerful tool to image the complex refractive index of a planar surface and to detect changes in a material's optical properties.

%\vspace{1em}
%OCIS Codes:

\end{abstract}

\maketitle

\vspace{1em}

\section{Introduction}
The study of reversible photodegradation of dye-doped polymers began in the early 2000's with the discovery of reversible photodegradation as measured with amplified spontaneous emission (ASE) from disperse orange 11 (DO11) doped into (poly)methyl-methacrylate (PMMA)\cite{howel02.01,howel04.01}.  Since then, reversible photodegradation has been found in Air Force 455 (AF455)\cite{Zhu07.01,Zhu07.02,Desau09.01}, rhodamine B and pyrromethene\cite{peng98.01}, as well as several anthraquinone derivatives\cite{Anderson11.02}.  Several methods have been used to probe the effect including ASE\cite{howel02.01,howel04.01,embaye08.01,Ramini11.01,Ramini12.01}, absorption spectroscopy\cite{embaye08.01}, fluorescence\cite{peng98.01}, digital imaging microscopy\cite{Anderson11.01,Anderson11.02}, and two-photon fluorescence\cite{Zhu07.01,Zhu07.02}.  In an effort to expand our ability to measure and understand reversible photodegradation, we have developed a white light interferometric microscope (WLIM), which combines the absorption spectrometer's frequency resolution with digital imaging's spatial resolution.  Along with measuring a sample's absorption spectrum, the WLIM measures the change in the real part of the index of refraction due to photodegradation.

A WLIM utilizes a Michelson interferometer and a CCD detector to obtain spatial and spectral resolution.  The method of using a Michelson interferometer with a photodiode detector to obtain spectral information in the visible spectrum is well known.  Using a point detector, Michelson interferometers have been used to measure the complex index of refraction in glass\cite{Dennis01.01}, gases\cite{Chamberlain69.01}, and liquids\cite{Engen98.01,Honjik73.01,Naganuma90.01}.  Further, it has been used with some modifications in astronomy research\cite{Hearn99.01} and the study of surfaces\cite{Deck03.01}.  The idea of using a CCD as a detector for interferometry was demonstrated previously by Pisani \textit{et. al.}; they used a Fabry-Perot interferometer, which measured only the imaginary part of the index of refraction\cite{Pisani09.01}.  Our WLIM is designed to spatially and spectrally resolve the photodamage induced change in the complex index of refraction of dye-doped polymer thin films.

\section{Method}

\subsection{Apparatus}
\begin{figure}[h!]
\centering
\includegraphics{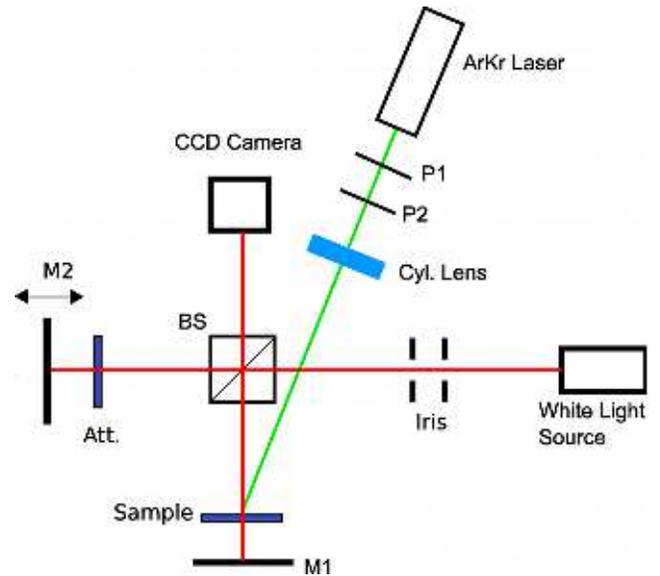}
\caption{White light interferometer setup.  M1 is the mirror in the stationary arm with the damaged sample; M2 is the moving mirror on a piezo stage; the reference arm has a balancing attenuator; BS is the cubic beam splitter; and P1 and P2 are polarizers used to control intensity. }
\label{Fig:Setup}
\end{figure}

The WLIM (see Figure \ref{Fig:Setup}) consists of a Thorlabs solid state light source (HPLS-30-03), a Michelson interferometer, and an Edmunds Optics monochrome CCD (EO 0813M).  The Michelson interferometer is made up of a UVFS uncoated non-polarizing cubic beam splitter and two uncoated UVFS mirrors, one of which is mounted on a Thorlabs piezo stage (NF5DP20S) with a piezo translation range of $20\mu$m.  The piezo stage is controlled by a closed feedback loop controller from Thorlabs (BPZ001), allowing for nanometer translation precision.

Photodegradation of samples is induced using an ArKr CW laser focused with a cylindrical lens, with two polarizers providing intensity control.  Control and data acquisition by computer uses the LabVIEW 2011 full development system, and the data is analyzed using custom procedures in Igor Pro.

\subsection{Interferometer Alignment Procedure}
A Michelson interferometer easily produces interference fringes when using a monochromatic laser with a coherence length of several meters.  The process is far more complicated when utilizing white light, which typically has a coherence length of the order of 40-60 $\mu$m.  The system is aligned first by optimizing the bulls-eye pattern from a HeNe laser beam that is collinear with the collimated white light. Then, a differential micrometer is used to set the two arms of the interferometer within 1mm of each other, at which point we scan the micrometer in intervals of approximately 30 $\mu$m in order to find the zero path length difference, where a series of dark and light fringes form.  The mirror alignment is subsequently adjusted so that the dark and light fringes change into a series of colorful fringes.  After several iterations, a centered white light interference pattern forms.  The ideal pattern is a bullseye shape with colorful fringes.  Since the optics have a minimum flatness of $\lambda/5$ due to the beam splitter, the pattern produced is oval in shape.

\subsection{Interferometer Theory}
The WLIM produces an intensity as a function of path length difference, $I(x)$, for each pixel of the camera, which can then be converted into the spectral intensity, $I(k_0)$, using a Fourier transform,

\begin{equation}
I(k_0)=\int^{\infty}_{-\infty}I(x)e^{-ik_0x}dx,
\end{equation}
where $k_0=\omega/c$ is the wavenumber in vacuum. The result of the Fourier transform is the interference intensity, which can be written as a function of the electric field in each arm of the interferometer as

\begin{equation}
I(k_0)\propto E^*_1(k_0)E_2(k_0)+E_1(k_0)E^*_2(k_0).
\label{Eqn:Iint}
\end{equation}
Given an incident electric field amplitude, $E_0(k_0)$, the electric field in each arm can be written as

\begin{equation}
E_i(k_0)=E_0(k_0)S_i(k_0)e^{i\Phi_i}
\label{Eqn:Efield}
\end{equation}
where $i=\{1,2\}$ denotes the arm, $S_i(k_0)$ is the spectral response of the optics in arm $i$, and $\Phi_i$ is the phase due to arm $i$.  Assuming $E_0(k_0)$ and $S_i(k_0)$ are real quantities and substituting Equation \ref{Eqn:Efield} into Equation \ref{Eqn:Iint} we find the interference intensity to be

\begin{equation}
I(k_0)\propto E_0(k_0)^2S_1(k_0)S_2(k_0)\exp\{i\Phi(k_0)\}+c.c. ,
\end{equation}
where $\Phi(k_0)=\Phi_2(k_0)-\Phi_1(k_0)$ and $c.c.$ denotes the complex conjugate.

\subsubsection{Empty interferometer}
For the empty interferometer the zero path length difference phase, $\Phi(x=0)$, in each arm can be written

\begin{eqnarray}
\Phi_1=2k_0y+\phi_1(k_0),
\\ \Phi_2=2k_0y+\phi_2(k_0),
\end{eqnarray}
where $2y$ is the balanced round trip arm length, and $\phi(k_0)$ is a phase introduced by the optics and deviations from the plane-wave approximation.  Combining the phases, we find that for the empty interferometer the phase difference, $\Phi(k_0)$, between the arms is

\begin{equation}
\Phi(k_0)=\phi_2(k_0)-\phi_1(k_0).
\end{equation}

\subsubsection{Samples in both interferometer arms}
Given the high optical density of dye-doped polymer samples and their relatively large index of refraction ($n\approx 1.5$ \cite{Yakuphanoglu08.01}), we must place nearly identical samples in each arm to maintain fringe contrast. ``Nearly identical'' means the samples' compositions are identical such that their complex index of refraction is the same, but their thickness and roughness may be different.  The zero path length difference phase of each arm is
\begin{eqnarray}
\Phi_1=2k_0(y-d_1-a_1)+2k_g(k_0)a_1+2\tilde{k}(k_0)d_1 \nonumber
\\+\psi_1(k_0)+\phi_1(k_0),
\end{eqnarray}
and
\begin{eqnarray}
\Phi_2=2k_0(y-d_2-a_2)+2k_g(k_0)a_2+2\tilde{k}(k_0)d_2 \nonumber
\\ +\psi_2(k_0)+\phi_2(k_0),
\end{eqnarray}
where $d_{1,2}$ is the sample thickness, $a_{1,2}$ is the glass substrate thickness, $k_g(k_0)$ is the real wavenumber of the glass where we assume the imaginary portion is negligible, $\tilde{k}(k_0)=k_0\tilde{n}(k_0)$ is the complex wavenumber of the dye-doped polymer, and $\psi_{1,2}(k_0)$ is a phase factor introduced due to the samples not being perfectly flat and aligned; $\phi_{1,2}$ comes from the empty interferometer phase. Combining phases and separating into the real, $\Phi'$, and imaginary, $\Phi''$, parts we find,

\begin{align}
\Phi'=2(d_2-d_1)[k'(k_0)-k_0]+2(a_2-a_1)\left[k_g(k_0)-k_0\right] \nonumber
\\ +\psi_1(k_0)-\psi_2(k_0)+\phi_2(k_0)-\phi_1(k_0), \label{Eqn:Rephase}
\end{align}
and
\begin{align}
\Phi''=2(d_2+d_1)k''(k_0). \label{Eqn:Imphase}
\end{align}
Using the definitions of the real and imaginary parts of $\tilde{k}(k_0)$,

\begin{eqnarray}
k'(k_0)&=&k_0n'(k_0),
\end{eqnarray}
and
\begin{eqnarray}
k''(k_0)&=&\frac{\alpha(k_0)}{2},
\end{eqnarray}
where $\alpha(k_0)$ is the absorbance per unit length, and $n'(k_0)$ is the real part of the index of refraction, we rewrite Equations \ref{Eqn:Rephase} and \ref{Eqn:Imphase} as

\begin{align}
\Phi'=2(d_2-d_1)k_0\left[n'(k_0)-1\right]+2(a_2-a_1)k_0\left[n_g(k_0)-1\right] \nonumber
\\ +\psi_1(k_0)-\psi_2(k_0)+\phi_2(k_0)-\phi_1(k_0),\label{Eqn:Rephase2}
\end{align}
and
\begin{align}
\Phi''=(d_2+d_1)\alpha(k_0).\label{Eqn:Imphase2}
\end{align}

\subsubsection{Differential degradation}
While the assumption that only the sample thickness varies is a good approximation for fresh samples, this assumption is weakened when one sample is damaged.  Letting $n'_0(k_0)$ and $n'_d(k_0)$ denote the undamaged and damaged index of refraction, respectively, $\alpha_u(k_0)$ and $\alpha_d(k_0)$ denote the undamaged and damaged absorbance per unit length, respectively, and letting only the sample in arm 1 to be damaged, we can rewrite Equations \ref{Eqn:Rephase2} and \ref{Eqn:Imphase2} as

\begin{align}
\Phi'=2k_0(d_1-d_2+n'_0(k_0)d_2-n'_d(k_0)d_1) \nonumber
\\ +2(a_2-a_1)k_0\left[n_g(k_0)-1\right] \nonumber
\\+\psi_1(k_0)-\psi_2(k_0)+\phi_2(k_0)-\phi_1(k_0),   \label{eqn:phasedeg}
\end{align}
and
\begin{align}
\Phi''=d_2\alpha_u(k_0)+d_1\alpha_d(k_0).
\end{align}
The difference between the undamaged and damaged phases is

\begin{eqnarray}
\Phi'_u(k_0)-\Phi'_d(k_0)=2k_0d_1[n_d(k_0)-n_0(k_0)],
\label{Eqn:dphase}
\end{eqnarray}
and
\begin{eqnarray}
\Phi''_u(k_0)-\Phi''_d(k_0)=d_1[\alpha_d(k_0)-\alpha_u(k_0)].
\end{eqnarray}

\subsection{Sample Preparation}
The samples used in this study are spin coated thin films of 1,4-Diamine-9,10-anthraquinone(1,4-DAAQ) doped into PMMA.  Samples are prepared as follows; PMMA and 1,4-DAAQ purchased from Sigma-Aldrich are dissolved into a solution of 33\% $\gamma$-Butyrolacetone and 67\% propylene glycol methyl ether acetate.  Maintaining a ratio of 15\% solids to 85\% solvents, we add the dye and polymer such that the concentration of dye in the polymer is 2.5g/l.  The solution is then stirred using a magnetic stirrer for 24 hours to ensure the dye and polymer fully dissolve.  Afterwards the solution is filtered through a 0.2$\mu$m ACRODISC filter to remove any remaining solids.

The glass substrates are prepared for spin coating by submerging plain glass slides in acetone to remove any residues from manufacturing, then placing the cleaned slides in deionized water and finally drying and storing them in a lint-free container to minimize contamination.  Once the dye-polymer solution is prepared, it is placed on a substrate and then spin coated at 1100 RPM for 90s.  The spin coated sample is then placed in an oven overnight to dry and to allow solvents to evaporate.  Once the sample has cooled, it is cut in half to form two nearly identical 2cm $\times$ 2cm squares.  When placing a sample in the interferometer for degradation, we use the two halves as a balanced pair in order to minimize differences between the sample arm and the attenuating arm of the interferometer.

\subsection{Experimental Procedure}
The experimental procedure is as follows.  The white light source warms up for one hour before taking data to minimize fluctuations in the light source.  Next, a reference interferogram is produced using the empty interferometer and a translation step size of 20nm, for a total number of 1000 steps.  At each step, the average of ten images is used in order to minimize noise from the CCD detector.  Once all the images are taken, they are imported into Igor Pro, where a custom procedure finds the interferogram at each pixel and then takes the FFT in order to find the phase and magnitude at each pixel.

Once the reference interferogram is taken, the sample and attenuator are mounted and the interferometer is realigned to compensate for wavefront distortion due to the samples not being perfectly homogeneous and flat.  This adjustment is found to effect the measured phase but not the magnitude.  Given that we are typically only concerned with differences due to photodegradation, the absolute phase is unimportant.  The pristine sample's interferogram is measured.  Subsequently, the sample is damaged using a pump laser with an average intensity of 60W/cm$^2$ for two hours, then another interferogram is taken. Finally, the sample is removed and another reference interferogram is taken to ensure the probe light has not drifted from its original intensity.

Using the magnitude of the interferogram at each pixel without the samples, with the pristine sample, and after damaging, the absorbance before and after photodegradation is determined at each pixel.  Given that an individual pixel is noisy, an average over adjacent pixels is performed to find a binned absorbance value.  The same averaging is applied to the phase data.  Since each pixel corresponds to a different point along the pump profile, spatial variations allow us to compare the change in absorbance and change in phase for different pump intensities.

\section{Results}

\subsection{Absorbance}
\begin{figure}[h!]
\centering
\includegraphics{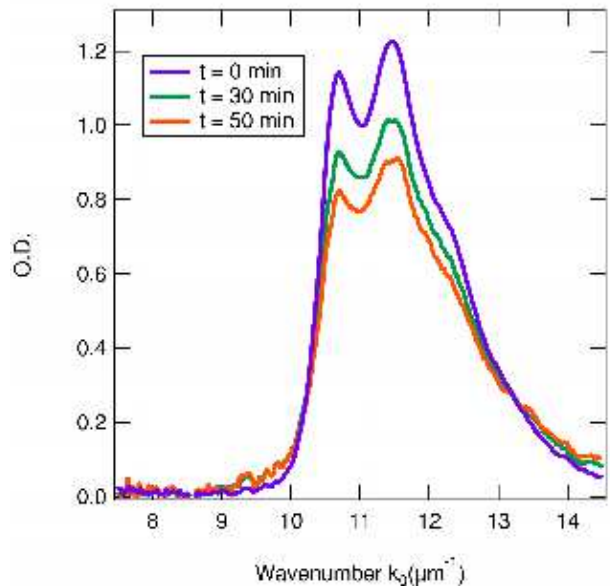}
\caption{Absorbance data as a function of time during photodegradation as measured with an Ocean Optics spectrometer}
\label{Fig:SpecAbs}
\end{figure}

\begin{figure}[h!]
\centering
\includegraphics{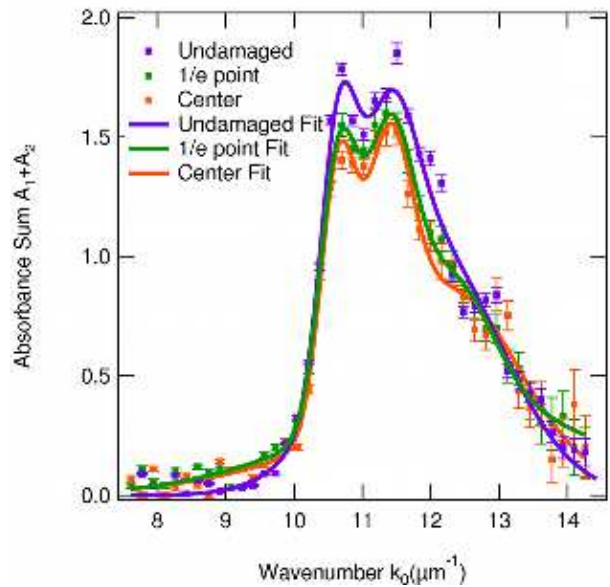}
\caption{Absorbance measured with the WLIM are shown as points.  Fits to a 4-peak Gaussian are shown as solid curves, which represent the absorption spectrum. Data were taken for an undamaged sample and for a damaged sample at the $1/e$ point of the burn profile and at the center of the burn profile.}
\label{Fig:WLIMabs}
\end{figure}

Given that one of the primary goals of developing the WLIM was to measure absorbance as a function of position and therefore intensity, the method is tested for each pixel by comparing the absorbance data for photodegradation measured with the WLIM to results found with an Ocean Optics spectrometer. The spectrometer accommodates thicker samples because of its greater dynamic range.  This difference in samples means the absolute numbers will be different between our experiments, but the spectral shapes remain the same.  Figure \ref{Fig:SpecAbs} shows spectrometer data at three times during photodegradation corresponding to different pump doses.  Figure \ref{Fig:WLIMabs} shows the sum of the absorbance in each interferometer arm, $A_1+A_2$,  as measured by the WLIM for the initial undamaged sample as well as the absorbance after decay at the burn center and $1/e$ position of the burn.  A 4-peak Gaussian fit of the WLIM data represents the absorption spectrum and confirms the WLIM result is consistent with spectrometer results.

\begin{figure}
\centering
\includegraphics{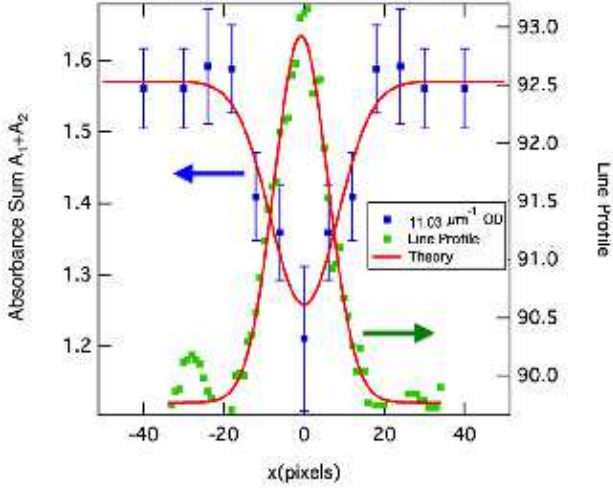}
\caption{Comparison of the WLIM optical density profile and the profile of a burn line determined from an imaging microscope.  The Gaussian width of both the WLIM measurement and the line profile are within experimental uncertainty.}
\label{Fig:spacecomp}
\end{figure}

For clarity, Figure \ref{Fig:WLIMabs} shows only the initial absorbance and two other points on the sample. Absorbance measurements were taken at many points along the burn line and averaged. Figure~\ref{Fig:spacecomp} shows the absorbance at $k_0 = 11.03 \mu$m$^{-1}$, the peak of the spectrum, as a function of pixel transverse to the line. Also shown is a profile of the burn line as imaged with an imaging microscope.

The Gaussian fit of both the absorbance sum and the line profile yield the same Gaussian widths within experimental uncertainty, showing the WLIM can spatially resolve the burn line.

\subsection{Phase}
\begin{figure}
\centering
\includegraphics{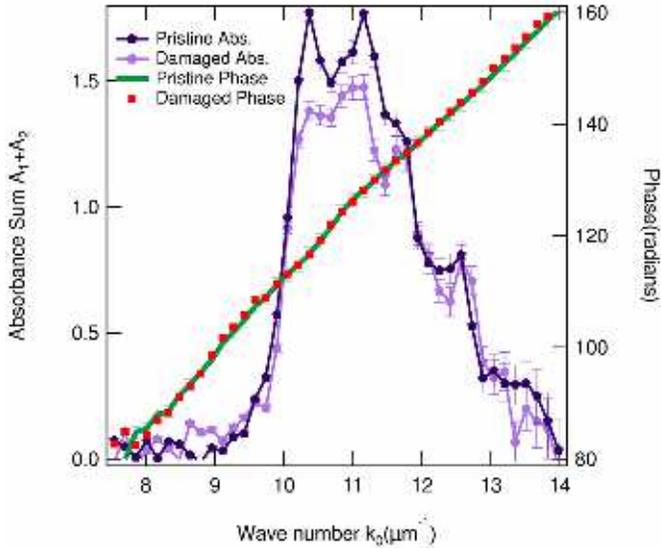}
\caption{Pristine and damaged (at center of burn) absorbance spectra as well as pristine and damaged phase as measured by the WLIM.}
\label{fig:HPdec}
\end{figure}

One of the benefits of interferometry as compared to ordinary spectroscopy is the ability to simultaneously measure both the real and imaginary parts of the index of refraction.  The real part of the index of refraction can be measured using the phase found from the Fourier Transform of an interferogram.  \textcolor{black}{ Figure \ref{fig:HPdec} shows the absorbance and phase for the pristine sample and the damaged sample at the burn center.  Upon inspection it is found that the change in phase due to photodegradation is small compared to the noise limit of the WLIM.  However, we find that calculating the optical pathlength difference (OPD) both before and after photodegradation allows us to resolve the change in refractive index.}

\textcolor{black}{To calculate the OPD, both before and after photodegradation, we begin with Equation \ref{eqn:phasedeg} and assume that $\psi_1=\psi_2=\phi_1=\phi_2=0$.  Doing so, Equation \ref{eqn:phasedeg} becomes}

\begin{align}
\textcolor{black}{\Phi'=2k_0(d_1-d_2+n'_0(k_0)d_2-n_d'(k_0)d_1)}\nonumber
\\ \textcolor{black}{+2k_0(a_2-a_1)[n_g(k_0)-1].}
\end{align}
\textcolor{black}{By measuring the phase due to the glass substrate alone, $\Phi_g'$, we can subtract out its contribution to find:}

\begin{align}
\textcolor{black}{\Phi'-\Phi_g'=2k_0(d_1-d_2+n'_0(k_0)d_2-n_d'(k_0)d_1).} \label{eqn:woop}
\end{align}
\textcolor{black}{From Equation \ref{eqn:woop} the OPD ($=\Phi/2k_0$) may be calculated as:}

\begin{align}
\textcolor{black}{\xi} &\textcolor{black}{=\frac{\Phi'-\Phi_g'}{2k_0},}
\\ &\textcolor{black}{=d_1-d_2+n'_0(k_0)d_2-n_d'(k_0)d_1,}  \label{eqn:OPD}
\end{align}
\textcolor{black}{where once again $n_0'$ is the undamaged refractive index and $n_d'$ is the damaged refractive index.}

\begin{figure}
\centering
\includegraphics{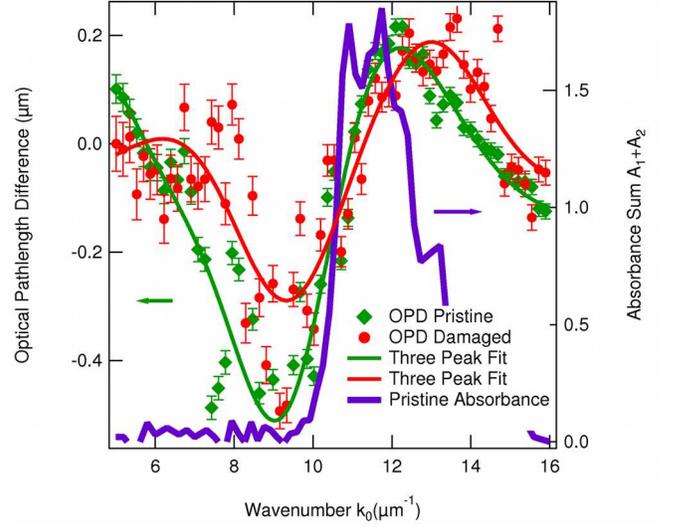}
\caption{\textcolor{black}{OPD for pristine sample (green points), after photodegradation (red points), and a three peak Gaussian fit (curves).  Pristine absorption spectrum is shown for comparison.}}
\label{fig:sn}
\end{figure}

\textcolor{black}{Figure \ref{fig:sn} shows the OPD, both for the pristine sample and in the region of peak damage, with fits to a three peak Gaussian function; the absorption spectrum for a pristine sample is included for comparison.  We find that the pristine OPD is consistent in both peak locations and widths with previous measurements of the refractive index of 1,4-DAAQ\cite{Yakuphanoglu08.01}. }

\begin{figure}
\centering
\includegraphics{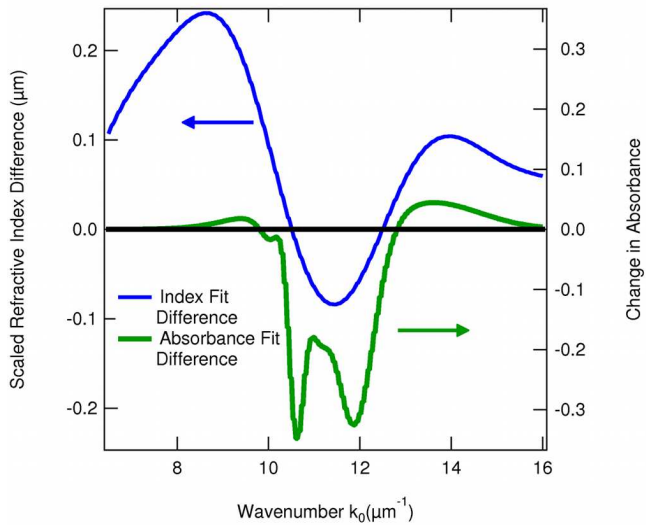}
\caption{\textcolor{black}{Scaled refractive index difference and absorbance change due to photodegradation.  Curves are calculated from fits to the WLIM data shown in Figure \ref{fig:sn}.}}
\label{fig:InRdiff}
\end{figure}

\textcolor{black}{Using Equation \ref{eqn:OPD} for the sample before and after degradation, we calculate the change in OPD, which gives the scaled refractive index difference:}

\begin{align}
\textcolor{black}{\xi_d-\xi_0=d_1(n_0'-n_d')}
\end{align}
\textcolor{black}{With multi-peak Gaussian fits of the OPD before and after degradation, Figure \ref{fig:InRdiff} shows the calculated scaled refractive index difference and the absorbance change due to photodegradation.  We find the change in refractive index is zero near the peak change in the absorption spectrum, which is consistent with a simple two-species two-level-molecule model of the refractive index as discussed in Appendix \ref{app}.  As such, the refractive index spectrum is consistent with absorbance measurements in suggesting one species converts into another during the photodegradation process.  To the lowest order of approximation, the polymer does not seem to be involved in degradation, though the vertical offset of the data may signal the contribution of the polymer, which has little dispersion in its refractive index in this wavelength range.}

\section{Conclusion}
A white light interferometric microscope can spatially resolve the change in the complex index of refraction due to photodegradation.  The spatially resolved absorbance measurements during decay are found to be consistent with spectrometer measurements.

 \textcolor{black}{We measured the pristine OPD of 1,4-DAAQ using the WLIM and found our results to be consistent with previous measurements\cite{Yakuphanoglu08.01}.  Additionally we measured the OPD after photodegradation and observed a change in the refractive index due to photo-induced damage, which is consistent with the decay process converting one species into another.}

\section{Acknowledgements}
 We would like to thank Wright Patterson Air Force Base and Air Force Office of Scientific Research (FA9550- 10-1-0286) for their continued support of this research.

\clearpage
\appendix*

\section{\textcolor{black}{Two-species two-level molecule model}}
\label{app}
\textcolor{black}{Embaye's simple model of the effect of reversible photodegradation on a sample's refractive index\cite{embaye08.01} assumes the system consists of two-level molecules of one species, which are converted into another two-level species via photodegradation.  For this system we can write the complex index of refraction as a function of frequency, $\omega$,}
\begin{equation}
\textcolor{black}{\tilde{n}(\omega)=\sqrt{1+m\tilde{\chi}_1+(1-m)\tilde{\chi}_2},} \label{eqn:ind}
\end{equation}
\textcolor{black}{where $m$ is the normalized population density of the undamaged molecules, $1-m$ is the density of damaged molecules, and $\tilde{\chi}_i$ is the complex susceptibility of the $i^{th}$ species given by}

\begin{align}
\textcolor{black}{\tilde{\chi}_j(\omega)=A_j \bigg(\frac{(\omega_{0,j}-\omega)T_{2,j}^2}{1+(\omega_{0,j}-\omega)^2T_{2,j}^2+\Omega_j^2T_{2,j}\tau_j }} \nonumber
\\ \textcolor{black}{+\frac{i\Omega_j T_{2,j}}{1+(\omega_{0,j}-\omega)^2T_{2,j}^2+\Omega_j^2T_{2,j}\tau_j }\bigg),} \label{eqn:chi}
\end{align}
\textcolor{black}{where $j=1$ for the undamaged species, $j=2$ for the damaged species, $A_j$ is an amplitude factor accounting for transition moments, $\hbar \omega_{0,j}$ is the transition energy, $\Omega_j$ is the Rabi frequency, $\tau_j$ is the natural lifetime of the excited state, and $T_{2,j}$ represents frequency broadening due to thermodynamics.}

\begin{figure}[h!]
\centering
\includegraphics{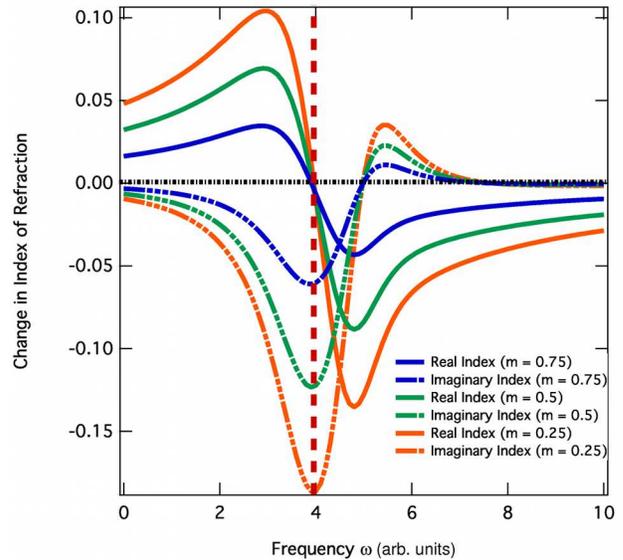}
\caption{\textcolor{black}{Change in the real and imaginary parts of the refractive index due to undamaged molecules degrading into the damaged species.  $m$ is the fraction of undamaged molecules.  The wavelength of peak change (vertical red dashed line) in the imaginary part corresponds to a zero-crossing point in the real part.}}
\label{fig:2p2l}
\end{figure}

\textcolor{black}{Substituting Equation \ref{eqn:chi} into Equation \ref{eqn:ind} and letting $\omega_{0,1}=5$ and $\omega_{0,2}=4.5$, Figure \ref{fig:2p2l} plots the change in both the real and imaginary parts of the refractive index for several population fractions of decay.  The peak change in the imaginary part of the index of refraction (absorbance) corresponds to a zero-crossing point of the real part of the refractive index.  This behavior is characteristic of of the conversion of one species into another, as is found during photodegradation.}

%\bibliographystyle{plain}
%\bibliography{PrimaryDatabase.bib}

\end{document}